# Can Cu(II) ions be doped into the crystal structure of potassium hydrogen tartrate?

Bikshandarkoil R Srinivasan[1] & H Remesh[2]
[1]Department of Chemistry, Goa University, Goa 403206, India Email: srini@unigoa.ac.in
[2]Department of Chemistry, Kerala University, Thiruvananthapuram 695581

**Abstract**

The differing binding preferences of the hydrogen tartrate ligand $(HC_4H_4O_6)^-$ namely $\mu_7$-octadentate mode for potassium ion and bidentate mode for cupric ion rules out the doping (incorporation) of any Cu(II) ion into the crystal structure of potassium hydrogen tartrate. Hence, the claim of growth of copper doped potassium hydrogen tartrate viz. $K_{0.96}Cu_{0.04}C_4H_5O_6$ by Mathivanan and Haris, *Indian J Pure App Phys* 51 (2013) 851-859 is untenable.

**Introduction**

The authors of a recent paper (title paper hereinafter) report on the growth of so called copper doped potassium hydrogen tartrate $K_{0.96}Cu_{0.04}(HC_4H_4O_6)$ **1** $((HC_4H_4O_6)^-$ is hydrogen tartrate) and iron doped potassium hydrogen tartrate $K_{0.96}Fe_{0.04}(HC_4H_4O_6)$ **2** by gel method[1]. Tartaric acid represented by the formula $(H_2C_4H_4O_6)$ exhibits chirality. It exists in nature as *L*-tartaric acid (dextrotartaric acid or $(+)^2$ form) or its enantiomer *D*-tartaric acid (laevotartaric or $(-)^2$ form). Mesotartaric acid and *DL*-tartaric acid ($\pm$ form) which is a 1:1 mixture of the *D*- and *L*-forms are optically inactive[3]. Although the authors have not stated in the title paper if an optically active or inactive form of tartaric acid was used for the growth of **1** or **2**, it can be assumed that an optically active form was possibly used in view of the comparison of the unit cell data with that of potassium hydrogen (+)-tartrate.

The study of tartaric acid and its metal salts especially potassium hydrogen (+)-tartrate has contributed to our understanding of the phenomenon of chirality and advancement of crystallography, details of which are reported by Derewenda[3]. Since the pioneering report of Bijovet et al[4] on the determination of absolute configuration of (+)-tartaric acid, several metal (+)-tartrates (and also $\pm$ tartrates) have been structurally characterized[5-13]. From the reported structures, it is inferred that the series of metal hydrogen (+)-tartrates having formula $[MH(C_4H_4O_6)]$ (M = K, Rb, Cs, NH$_4$) are isostructural and crystallize in the orthorhombic space group $P2_12_12_1$ with Z = $4$[5-9]. Although the hydrogen tartrates of $K^+$, $Rb^+$, $Cs^+$ etc. are crystallized from aqueous solution, they are anhydrous unlike the hydrogen tartrates of Li, Na, Cu, Co etc which contain one or more water molecules[10-13].

In view of the above mentioned literature data, the claim of doping (incorporation) of Cu(II) or Fe(II) ions into the structure of potassium hydrogen tartrate by Mathivanan and Haris[1] appeared quite unusual. Further, the charge imbalance of both the doped crystals for example in the formula $K_{0.96}Cu_{0.04}(HC_4H_4O_6)$ for the mononegative $(HC_4H_4O_6)^-$ ion the positive charges on both the metals add up to 1.04 (in view of the +2 state of Cu) indicated that the claim of doping is unreliable. Hence, the title paper reporting on growth of doped tartrate crystals **1** and **2** attracted our attention and was taken up for scrutiny to determine if such charge imbalanced metal tartrates can be crystallized.

**Can Cu(II) ions be doped into the crystal structure of potassium hydrogen tartrate?**

In recent work from our laboratory we have shown that known structural features are useful to determine which ion or molecule can be incorporated into the crystal structure of a compound[14, 15]. In the following we present a discussion of the known structures of metal hydrogen tartrates. In view of the isostructural nature of $[MH(C_4H_4O_6)]$ (M = K, Rb, Cs, $NH_4$) complexes, the $Rb^+$, $Cs^+$ $(NH_4)^+$ ions can be expected to be incorporated into the crystal structure of potassium hydrogen tartrate, as can be evidenced by the structural characterization of $K_{0.56}(NH_4)_{0.44}(C_4H_5O_6)$ and $K_{0.5}Rb_{0.5}(C_4H_5O_6)$ containing the racemic hydrogen tartrate ligand by Gelbrich et al[9]. These doped crystals isolated from aqueous solution[9], are perfectly charge balanced.

No examples of structurally characterized hydrogen tartrates containing Cu(I) or Fe(I) are reported in the Cambridge Structure Database (CSD) till date. In addition to the charge imbalance in the crystals of formula $K_{0.96}M_{0.04}(HC_4H_4O_6)$ (M=$Cu^{+2}$ or $Fe^{+2}$) which precludes the incorporation of bivalent metal ions like Cu(II) or Fe(II), the differing structural features of the hydrogen tartrates of $K^+$, and $Cu^{+2}$ shows a structural mismatch. This is explained by a comparison of the crystal the structures of potassium hydrogen tartrate (KHT), (formula = $[K(HC_4H_4O_6)]$) and diaquabis(hydrogen tartrato)copper(II) dihydrate (formula = $[Cu(HC_4H_4O_6)_2(H_2O)_2]\cdot 2H_2O$). The molecular formulae which differ considerably reveal that KHT is anhydrous while the Cu(II) salt contains four water molecules and two hydrogen tartrates per copper unlike KHT. The central K⁺ ion in KHT is eight coordinated while the Cu(II) ion is six coordinated (Fig. 1 and 2).

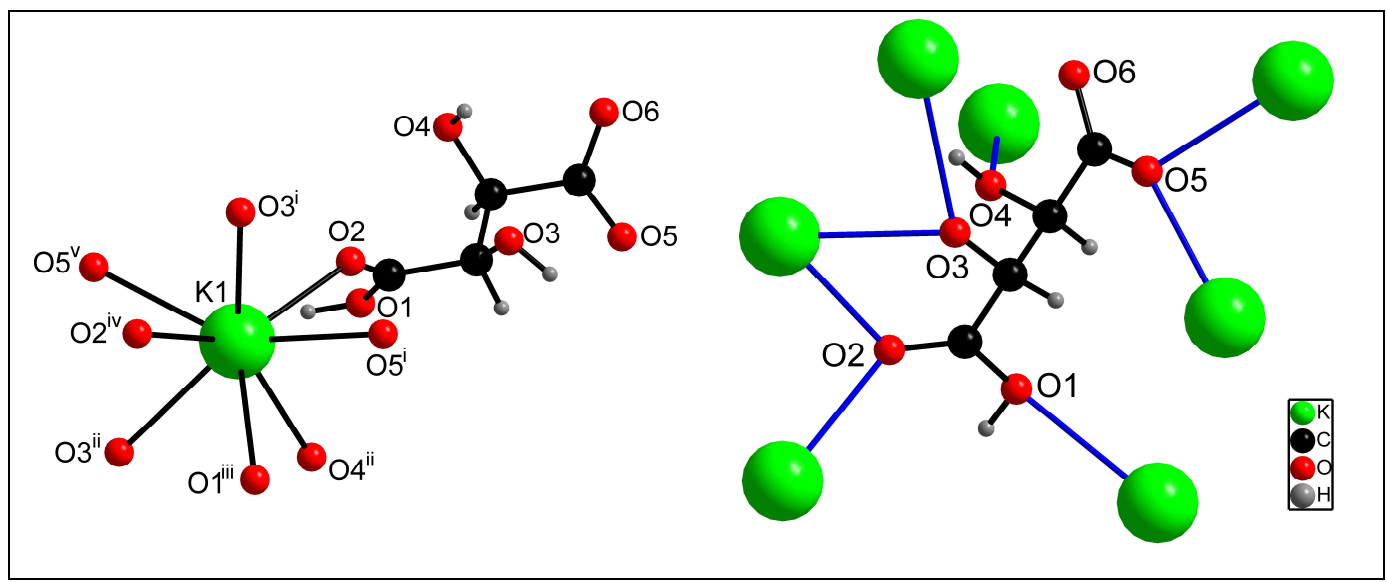

**Fig. 1** Crystal structure of [K(HC$_4$H$_4$O$_6$)] showing the eight coordination around K⁺ (left). Symmetry codes: i) 1.5-x, 1-y, 0.5+z  ii) 0.5+x, 0.5-y, -1-z  iii) 1.5-x, 1-y, 1.5+z  iv) 1-x, 0.5+y, -1.5-z  v) -0.5+x, 0.5-y, -1-z.  The μ$_7$-octadentate coordination mode of the hydrogen tartrate ligand (right). Figure is drawn using the CIF file in Ref. 6

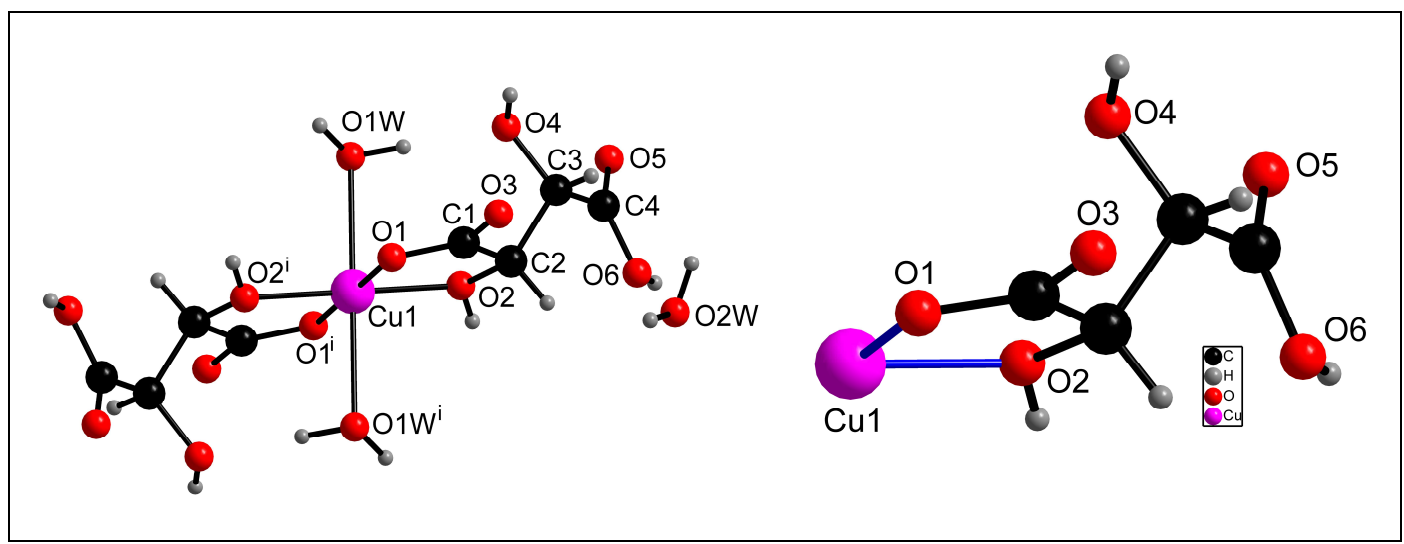

**Fig. 2** Crystal structure of [Cu(HC$_4$H$_4$O$_6$)$_2$(H$_2$O)$_2$] showing the six coordination around Cu²⁺. O1W and O2W are coordinated and lattice waters respectively. Symmetry code: i) 1-x, -y, -z (left).  The bidentate coordination of the hydrogen tartrate ligand (right).  Figure is drawn using the CIF file in Ref. 11.

The hydrogen tartrate ion functions as a bridging ligand and exhibits a $\mu_7$-octadentate coordination mode bridging seven different $K^+$ ions in the crystal structure, resulting in eight coordination around K and the formation of a three dimensional coordination polymer. In contrast, each of the hydrogen tartrate ligand functions as a bidentate ligand ($\eta_2$-coordination) thus accounting for four binding sites. The two monodentate aqua ($H_2O$) ligands complete the hexacoordination around Cu resulting in a discrete complex. Hence, the incorporation of any Cu(II) ions into the crystal structure of KHT can be ruled due to i) The differing stereochemistry and dimensionality of the tartrate complexes of K and Cu; ii) The differing binding preferences of the hydrogen tartrate ligand for K(I) and Cu(II) ions.

Regarding the claim of incorporation of Fe(II) into the structure of KHT we wish to mention that the authors have not taken into consideration that only $Fe^{+3}$ (not $Fe^{+2}$) is oxophilic and hence Fe(II) is not expected to bind to tartrate. In accordance with this no examples of tartrates of Fe(II) are known in the literature. We believe that the so called copper (or iron) doped potassium hydrogen tartrate crystals were grown and characterized not based on the known structural details of metal hydrogen tartrates and a proper interpretation of the experimental data, but based on an incorrect assumption that the use of specified quantity of Cu (or Fe) in the form of copper nitrate (or iron sulphate) along with KCl and tartaric acid will result in the formation of their desired copper or (iron) doped crystal. In order to justify their assumption, the authors have reported experimental % of K, Cu and Fe in pure KHT and copper doped KHT **1** and Fe doped KHT **2**. Although it is not clear as to how a weight % of 100 was obtained experimentally by the authors for pure KHT which contains C, H, and O in addition to K, the data can be considered as dubious because 100 % K is expected only for pure K metal (Table 1). We regret to point out that all values of theoretical % calculated by the authors are incorrect as the % of only K and Cu (or Fe) for **1** (or **2**) add up to 100% and does not take into account the other atoms namely C, H and O. The correct % of K and Cu (or Fe) for **1** (or **2**), C, H and O for the proposed formula (if indeed such compounds can exist) shown in Table 1 unambiguously indicate that no incorporation of Cu (or Fe) into the crystal structure of potassium hydrogen tartrate has taken place as claimed by the authors. Based on the reported experimental metal %, **1** and **2** can at best be declared as improperly characterized crystals but not as copper (or iron) doped potassium hydrogen tartrate crystals.

**Table 1** Theoretical weight % of K, Cu and Fe for pure potassium hydrogen (+)-tartrate (KHT) and so called copper (or Fe) doped KHT based on molecular formula

| Name | Formula | M.W. | % K | % dopant | % C | % H | % O | Total % |
|---|---|---|---|---|---|---|---|---|
| Pure tartaric acid | $C_4H_5O_6$ | 150.09 | --- | --- | 32.01 | 4.03 | 63.96 | 100 |
| Pure potassium | K | 39.10 | 100 | --- | --- | --- | --- | 100 |
| Pure KHT[#] | $KC_4H_5O_6$ | 188.18 | 20.78 **(100)** | --- **(0)** | 25.53 | 2.68 | 51.01 | 100 |
| so called Cu doped KHT[#] | $K_{0.96}Cu_{0.04}C_4H_5O_6$ | 189.16 | 19.84 **(92.56)** | 1.34 **(7.44)** | 25.40 | 2.67 | 50.75 | 100 |
| so called Fe doped KHT[#] | $K_{0.96}Fe_{0.04}C_4H_5O_6$ | 188.85 | 19.88 **(95.22)** | 1.18 **(4.78)** | 25.44 | 2.67 | 50.83 | 100 |

*Values in **bold** (in bracket) are the experimental values reported for pure KHT, so called Cu doped KHT and so called Fe doped KHT in the title paper. [#] No data reported for C, H and O.

We regret to mention that reporting a magnetic moment of 2.525 BM for pure $KH(C_4H_4O_6)$ cannot be correct since it is a well-known diamagnetic solid in view of the closed shell electronic configuration of the $K^+$ ion. The dubious nature of the magnetic data can be evidenced from the reported moments for the so called doped crystals which are less than the pure system in spite of addition of paramagnetic Cu(II) or Fe(II) ions. In view of the fundamental errors namely violation of charge neutrality and the structural mismatch of the hydrogen tartrates of K(I) and Cu(II), the other questionable results in the title paper do not merit any discussion.

In summary, we have shown that the structural features of potassium hydrogen (+)-tartrate do not permit the doping or incorporation of any Cu(II) (or Fe(II)) ions into its crystal structure.

**Acknowledgements**

HR thanks the Indian Academy of Sciences, Bangalore for financial support under the Summer Research Fellowship-2014 programme, to carry out research work at Goa University.